\documentclass[a4paper,twocolumn,prl,showpacs,showkeywords,aps,nofootinbib]{revtex4}
\usepackage{graphicx}
\usepackage{amsmath}
\usepackage{bm}
\newcommand{\beq}{\begin{equation}}
\newcommand{\eeq}{\end{equation}}
\newcommand{\beqa}{\begin{eqnarray}}
\newcommand{\eeqa}{\end{eqnarray}}
\def\ra{\rangle}
\def\la{\langle}

\begin{document}

\title{Atom cooling by non-adiabatic expansion}

\author{Xi Chen$^{1,2}$}

\author{J. G. Muga$^{1}$}

\author{A. del Campo$^{3,4}$}

\author{A. Ruschhaupt$^{5}$}

\affiliation{$^{1}$Departamento de Qu\'{\i}mica-F\'{\i}sica,
UPV-EHU, Apdo 644, 48080 Bilbao, Spain}

\affiliation{$^{2}$Department of Physics, Shanghai University,
200444 Shanghai, P. R. China}

\affiliation{$^{3}$Institute for Mathematical Sciences, Imperial College London, 53 Prince's Gate, SW7 2PG London, UK}
 
\affiliation{$^{4}$QOLS, Blackett Laboratory, Imperial College London, Prince Consort Road, SW7 2BW London, UK}

\affiliation{$^{5}$Institut f\"ur Theoretische Physik, Leibniz
Universit\"{a}t Hannover, Appelstra$\beta$e 2, 30167 Hannover,
Germany}

\begin{abstract}
Motivated by the recent discovery that a reflecting wall moving 
with a square-root in time trajectory behaves as a universal stopper of classical particles
regardless of their initial velocities,    
we compare linear in time and square-root in time expansions of a  box
to achieve efficient atom cooling. For the quantum single-atom wavefunctions
studied the square-root in time expansion presents 
important advantages: asymptotically it leads to zero average energy
whereas any linear in time (constant box-wall velocity) expansion leaves 
a non-zero residual energy, except in the limit of an infinitely slow expansion. 
For finite final times and box lengths we set a number of bounds 
and cooling principles which again confirm the superior performance 
of the square-root in time expansion, even more clearly for increasing excitation of the initial state. Breakdown of adiabaticity is generally
fatal for cooling with the linear expansion but not so with the square-root expansion.   
 


\end{abstract}

\maketitle

\begin{figure}[t]
\begin{center}
(a)\hspace*{0.5cm}\includegraphics[width=0.6\linewidth]{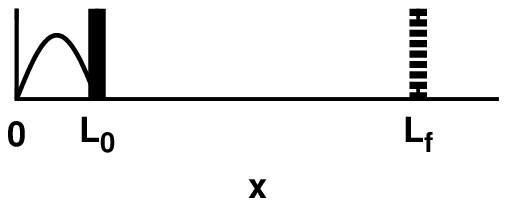}

(b)\includegraphics[width=0.7\linewidth]{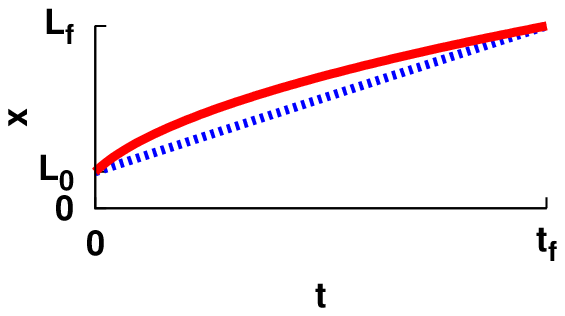}
\end{center}
\caption{\label{fig1} (Color online) (a) Scheme of atom cooling
by expanding a box. (b) The two right-wall trajectories considered in this work: 
linear mirror (dotted blue line) and
square-root mirror (solid red line), see eq. (\ref{trajectory}).}
\end{figure}

Cooling and trapping of atoms has been a central theme in physics
for over 30 years, because of applications in
precision measurements and fundamental physics
\cite{Metcalf}.
Stopping particles released from an oven or source continuously or in pulses  
is frequently one of the steps in a cooling process
\cite{Mark1,Mark2,David}. In this regard,
a relevant and surprising recent discovery is that a reflecting potential (or
``mirror'' hereafter) moving with a square-root in time (square-root for short) trajectory
stops all classical particles released from a point source irrespective of
their initial velocity \cite{Schmidt}.
This is in clear contrast with a linear in time (linear for short) mirror
trajectory which only stops particles with one specific velocity. Motivated by
this result, and by the similar behavior found for quantum wavepackets
\cite{Schmidt},
we investigate here the effect of
expanding a square box with linear or square-root wall trajectories on the average
energy of a confined atom, see Fig. \ref{fig1}.
This is a non-trivial extension of the one-wall configuration since the
confinement may induce multiple atom-wall collisions.

A given cooling objective may be achieved with an ``adiabatic'' expansion, see
\cite{adiab1,adiab2} as a sample of earlier and recent experiments, in which
the confining walls move slowly keeping the populations of instantaneous
levels constant.\footnote{This is the standard definition of ``adiabatic'' in
  quantum mechanics, in contrast to the thermodynamical definition associated
  with no heat exchange. For a connection see \cite{Pol}.} 
Adiabatic expansions constitute a traditional phase-space-conserving cooling
method. Note that cooling is frequently associated with phase-space
compression but in fact compression is only required
for specific purposes, most notably to achieve Bose-Einstein condensates.  
For many applications in interferometry and metrology, and in particular in time-frequency metrology, an ultracold gas is by now preferable to condensates, to avoid the perturbing effects of interactions,
or phase separation phenomena \cite{Band}. Evaporative cooling is moreover a highly inefficient process based on losing the fastest atoms whereas a cooling expansion may retain all initial atoms. The linear expansion
could lead to vanishing energy in an
infinitely slow expansion, but ``time'' is in general a scarce commodity in
the laboratory. An experiment with cold atoms is limited by the finite time
in which the atoms remain trapped due, for example, to three-body losses.
Moreover for applications such as pulse formation in atomic clocks, 
it is desirable to cool the atoms, without forming a condensate,  
in a short time. This increases the flux  
that crosses the Ramsey fields, but it also goes against 
adiabaticity. 
In this work we shall provide a way out: 
finite-time expansions achieving efficient cooling are possible by 
using fast, square-root wall trajectories that do not obey at all the
adiabaticity criterion.             

%

A number of results on expanding and contracting boxes are known 
in the context of chaotic dynamics and the Fermi-Ulam model \cite{Fermi}, which consists of  
a bouncing ball between a fixed wall and a periodically moving wall
$L(t)$. A related question after the 
pioneering work of Doescher and Rice \cite{Rice}
is to find functions $L(t)$, not necessarily periodic, so that the corresponding Schr\"odinger equation can be solved exactly \cite{Berry,Pinder,Makowski,Luz,Dodonov}. In particular, Berry and Klein \cite{Berry}
extended the analysis beyond square boxes to a more general family of
expanding fields scaling with $L(t)$.  
They found explicit solutions for 
$L(t)=(at^2+2bt+c)^{1/2}$ keeping the coefficient
combination $ac-b^2$ constant \cite{Berry}. The linear expansion corresponds to making this constant zero, whereas $a=0$ is the signature of a square-root expansion. Under these conditions, 
the  propagator can be explicitly written in terms of known ``expanding
modes'' $\phi_n(t)$,
see e.g. \cite{Makowski} for details.

\begin{figure}[]
\begin{center}
(a)\scalebox{0.37}[0.37]{\includegraphics{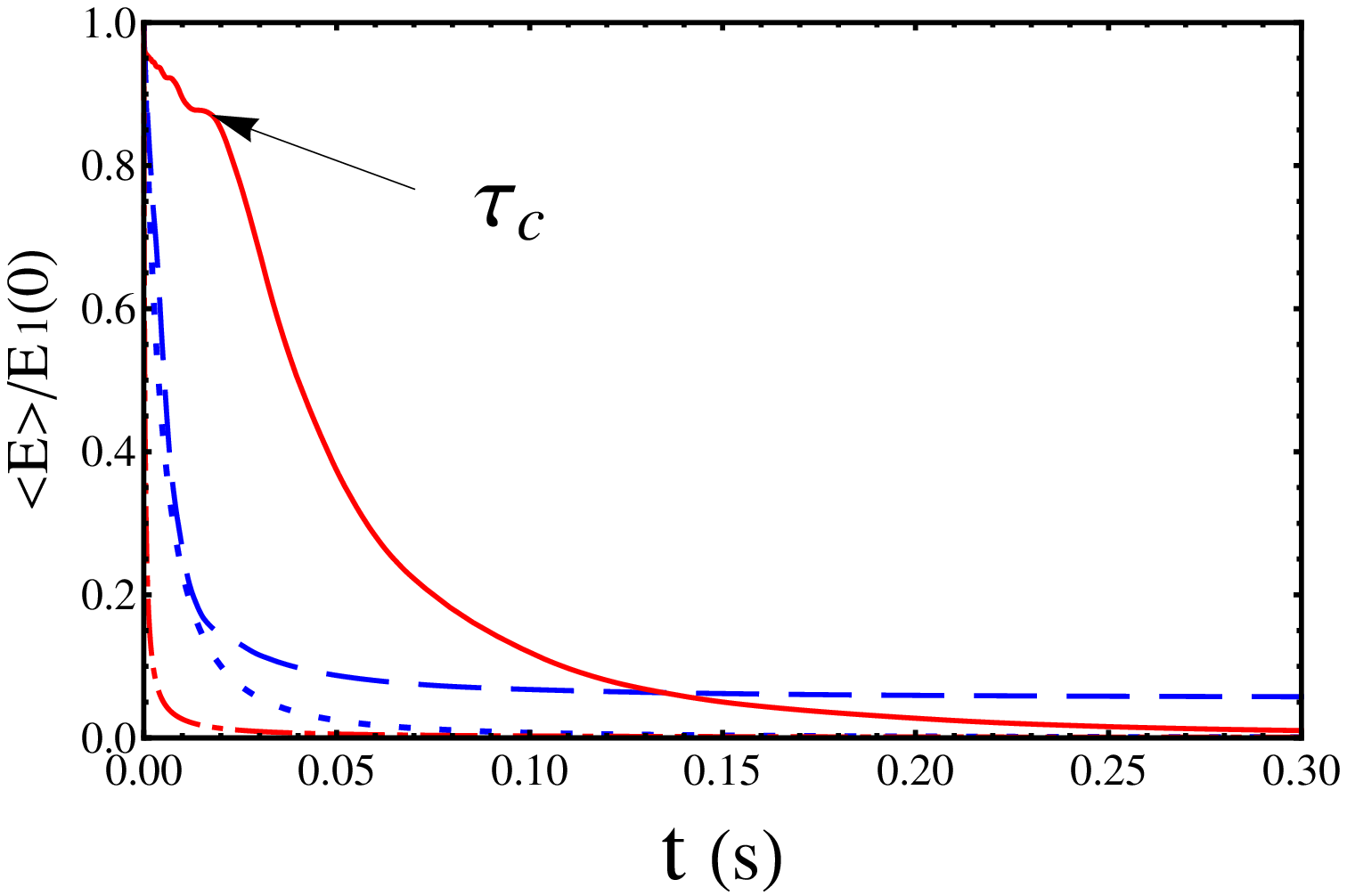}}

(b)\scalebox{0.4}[0.4]{\includegraphics{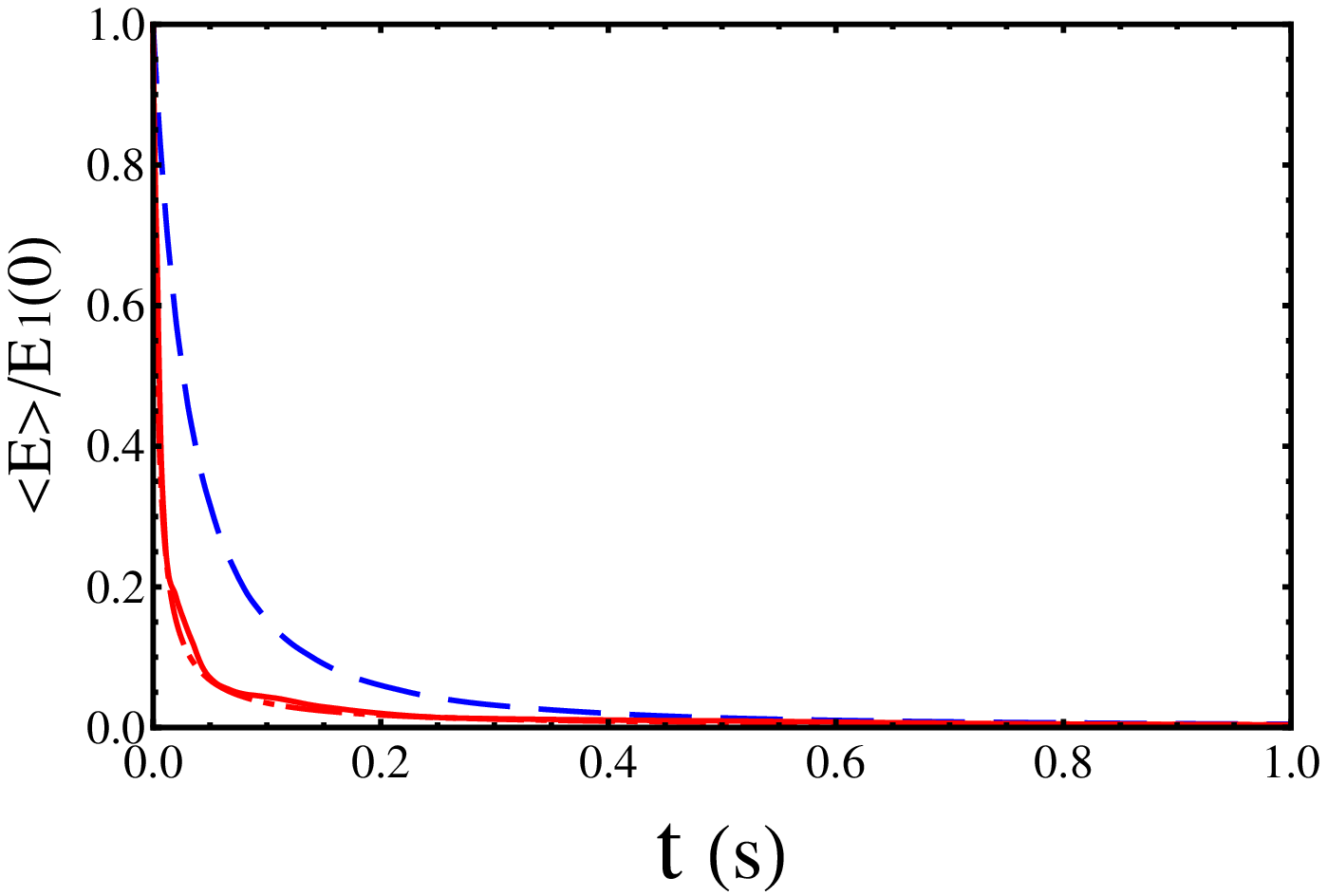}}
\end{center}
\caption{\label{fig2} (Color online) Dependence of the
average energy $\langle E \rangle$ on the time $t$:
square-root mirror (solid red line), linear mirror (dashed blue line);
adiabatic energies $E_{1}(t)$:
square-root mirror (dot-dashed red line), linear mirror (dotted blue line);
(a) $t_f =0.3$ s, $L_f = 100$ $\mu$m,
(b) $t_f =1 s$,  $L_f = 50$ $\mu$m,
in this figure part the adiabatic energies for the linear mirror
are indistinguishable from
the corresponding average energy in the scale of the figure;  
$L_0=3$ $\mu$m, initial state: ground state ($n=1$);
in all figures we use the mass of $^{87}$Rb atoms.}  
\end{figure}

For our purpose here the main result of Berry and Klein \cite{Berry} is  
that the energy of any expanding mode, $\la \phi_n(t)|H|\phi_n(t)\ra$ goes to zero 
at large time for $a=0$ (square-root expansion), but in general it tends to a constant 
proportional to $a$. If we note, in addition, that non-diagonal elements 
of the Hamiltonian matrix in this basis tend to zero asymptotically,
the conclusion is that  
a square-root expansion produces at long time a state with vanishing energy, whereas for a
linear one there will remain a non-zero residual energy. In the 
expanding box these are rather intuitive results, since one expects that for any finite constant-velocity expansion, there will be low atomic velocity components that cannot interact with the moving wall, whereas the 
square-root mirror keeps slowing down and interacts eventually with all components. 
This may be seen in the final energy values of Fig. \ref{fig2}a for rubidium-87, 
which we shall comment in more detail later on.   

%

Similarly to the time limitations mentioned before,
space is also bounded in practice. It is thus desirable to 
examine the effect of expansions for given final box length $L_f$ and time
$t_f$.

Consider the atoms in a quantum box with a moving hard wall as shown
in Fig. \ref{fig1}. The two different trajectories for the
``moving mirrors" are
\begin{eqnarray}
 L(t)= \left\{
\begin{array}{ll}
L_0 + [(L_f-L_0)/t_f] t &~ (\mbox{linear mirror})\hspace*{-0.8cm} \\[0.2cm]
\sqrt{L^2_0 + [(L^2_f-L^2_0)/t_f]t} &~ (\mbox{square-root mirror})\hspace*{-0.8cm}
\end{array}
\right.\nonumber\\
\label{trajectory}
\end{eqnarray}
where $L_0$ is the initial width of the box. 
It is convenient to expand the wave
function at an arbitrary time $t$ in the ``instantaneous'' basis 
$\{u_n(t)\}_{n=1,2,...}$ \cite{Schiff}, as
\begin{equation}
\label{wave function}
\psi (x, t)= \sum_{n=1}^\infty a_n(t) u_n(x,t) \exp\left[(i
\hbar)^{-1} \int^{t}_{0} E_n(t') dt'\right],
\end{equation}
where the instantaneous eigenstate of the box Hamiltonian  
is $u_n(x,t)=\sqrt{\frac{2}{L(t)}} \sin\left[\frac{n \pi x}{L(t)}\right]$, 
and the coefficients $a_n(t)$ are determined by solving the system of differential
equations that results from substituting $\psi$
into the Schr\"odinger equation.  

The average energy can be easily calculated as
\begin{equation}
\label{average energy} \langle E(t) \rangle = \sum_{n=1}^\infty p_n(t) E_n(t), 
\end{equation}
where $p_n(t)=|a_n(t)|^2$ is the population of the instantaneous eigenstate $n$, 
and $E_n(t)=\frac{n^2 \pi^2 \hbar^2 }{2 m L^2(t)}$.  
If the cooling objective is to diminish $\la E \ra$ up to a given value 
or to a fraction of its initial value for finite $t_f$ and $L_f$, a number of principles and bounds that govern the expansion process provide
useful guidance. We shall first note some general principles and then 
specific bounds that underline the usefulness of the square-root versus the linear mirror:       

{\it (i) General principles.} 
First, it is clear that a sudden expansion is useless. If the mirror moves too
fast the particle expands 
freely and the final energy $E_f=\la E(t_f)\ra$ will be equal to the initial energy,
$E_{f}= \la E(0)\ra$.

Moreover, the energy bound 
\begin{equation}
\langle E(t) \rangle \geq  E_1 (t) = \frac{\pi^2 \hbar^2}{2 m
L^2(t)}
\label{ebound}
\end{equation}
is always fulfilled. As an application, it sets a bound on the minimal final length for
a given cooling objective,   
\begin{equation}
L_f \geq  \frac{\pi \hbar}{\sqrt{2m E_f}}.
\end{equation}
For example, if a $1/100$ energy reduction is required starting from the ground state, the
final length of the box should satisfy $L_f \geq 10 L_0$.

Last but not least, the microscopic ``minimal work principle'' \cite{Allahverdyan}
states that  
the average energy satisfies $\la E\ra\geq E_{adiab}$,
where the ``adiabatic energy'' is 
$E_{adiab}(t)=\sum_n p_n(0)E_n(t)$,  provided the initial state is passive
(diagonal density matrix with   
$p_n \geq p_{n+1}$) and there are no level crossings. 
These conditions are satisfied for an 
expanding box and for the ground state as initial state.
An exception to the principle for a non-passive, excited initial state,
will be commented later on.   

\begin{figure}[]
\begin{center}
(a)\includegraphics[width=0.8\linewidth]{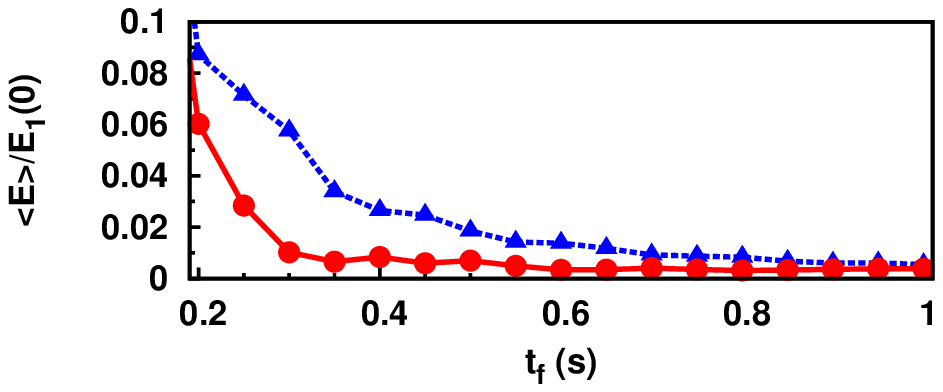}

(b)\includegraphics[width=0.8\linewidth]{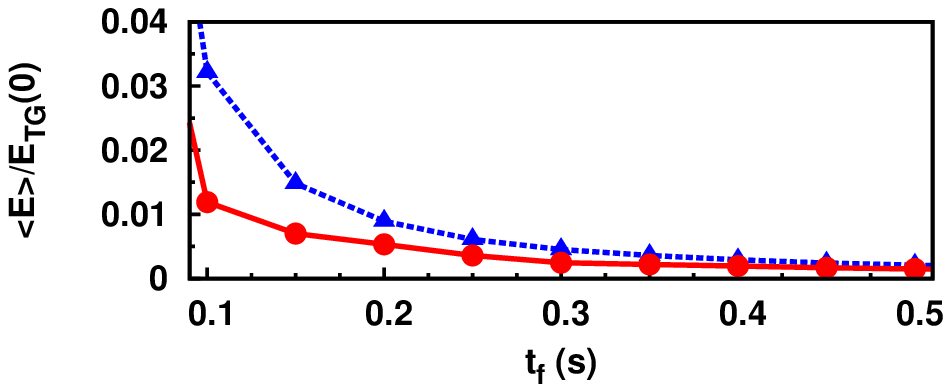}

\caption{\label{fig3} (Color online) Dependence of the final average
energy $E_f$ on the final time $t_f$:
square-root mirror (red circles connected with a solid line),
linear mirror (blue triangles connected with a dashed line);
(a) initial state: single-particle ground state ($n=1$);
(b) initial state: Tonks-Girardeau gas with $5$ particles,
$E_{TG}(0)$ is the energy of the
initial ground state of the Tonks-Girardeau gas;
$L_0=3$ $\mu$m, $L_f=100$ $\mu$m.}
\end{center}
\end{figure}

{\it (ii) Stoppable and unstoppable velocities.}
Classically, a necessary condition for stopping is that the particle
has reached the mirror before $t_f$. This implies a (classical) minimal
stoppable velocity for the linear mirror as well as the square-root mirror
(see also Fig. \ref{fig1}b), 
\begin{eqnarray}
\label{minimal velocity} v_{min}= L_f/t_f, 
\end{eqnarray}
which should be small compared to the typical velocities of the initial state.
For the ground state, and using the quasi-velocity $v_1=\pi\hbar/(mL_0)$
for an estimate, this gives the condition  
\beq
\label{condstop}
\frac{v_{min}}{v_1}=\frac{m L_f L_0}{t_f\pi\hbar}\ll 1
\Leftrightarrow \frac{m L_f L_0}{\pi\hbar} \ll t_f,
\eeq
with different implications for the two cases:
the square-root mirror stops
(for $L_0/L_f \to 0$) all classical particles
moving with velocity greater than $v_{min}$, whereas
this is not true for the linear mirror.  

%
\begin{figure}[]
\begin{center}
(a)\scalebox{0.33}[0.33]{\includegraphics{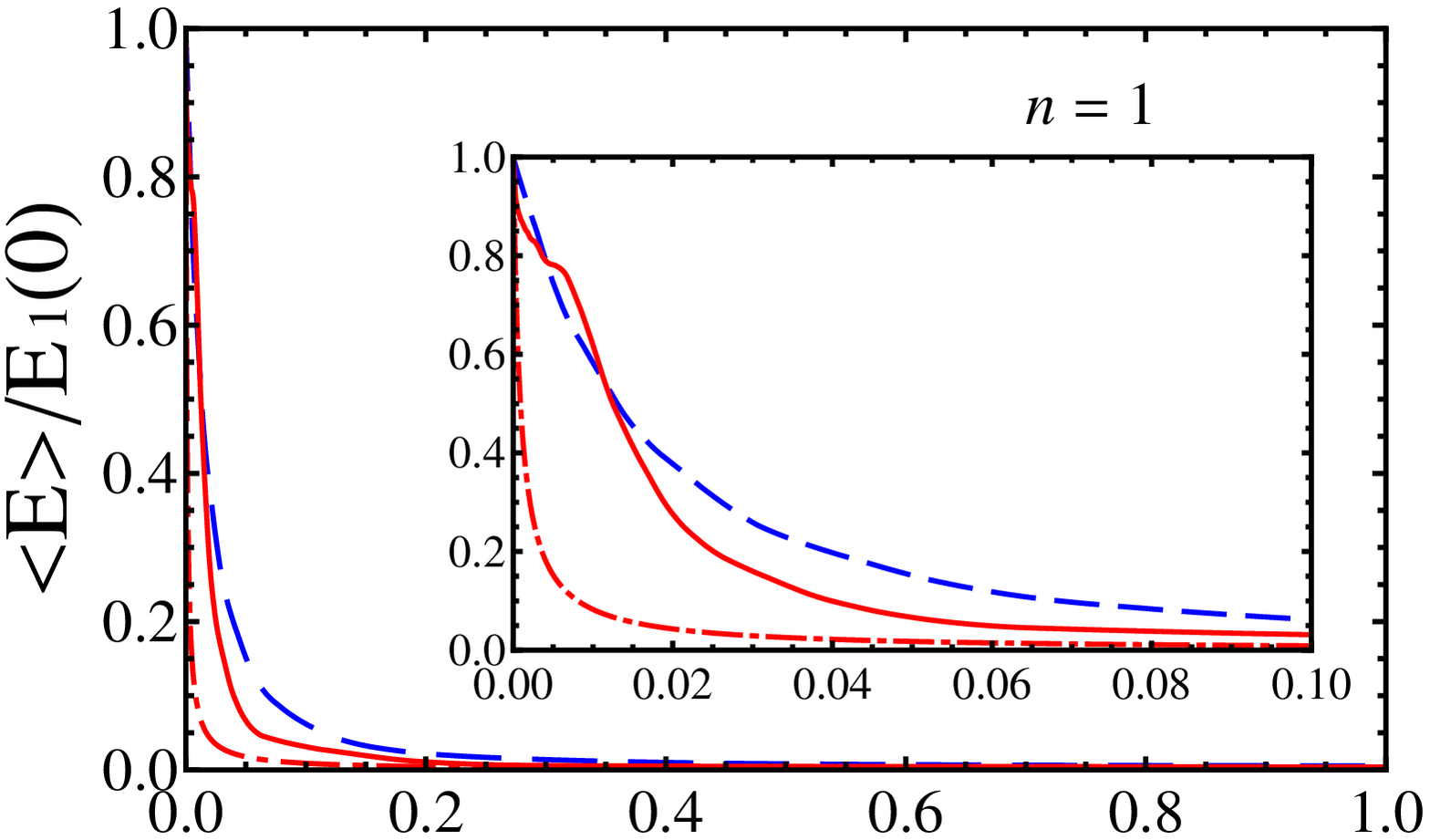}}

(b)\scalebox{0.33}[0.33]{\includegraphics{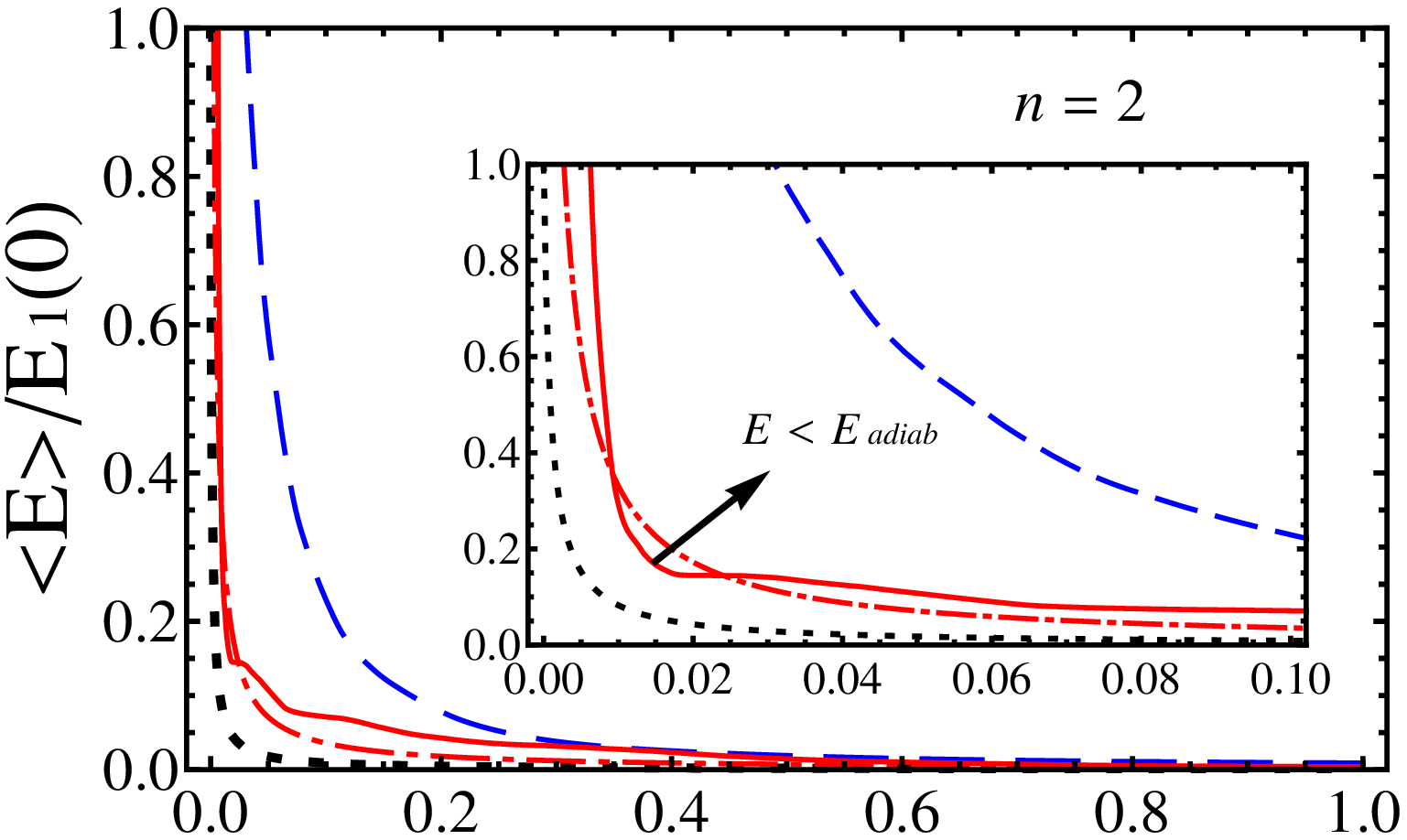}}

(c)\scalebox{0.33}[0.33]{\includegraphics{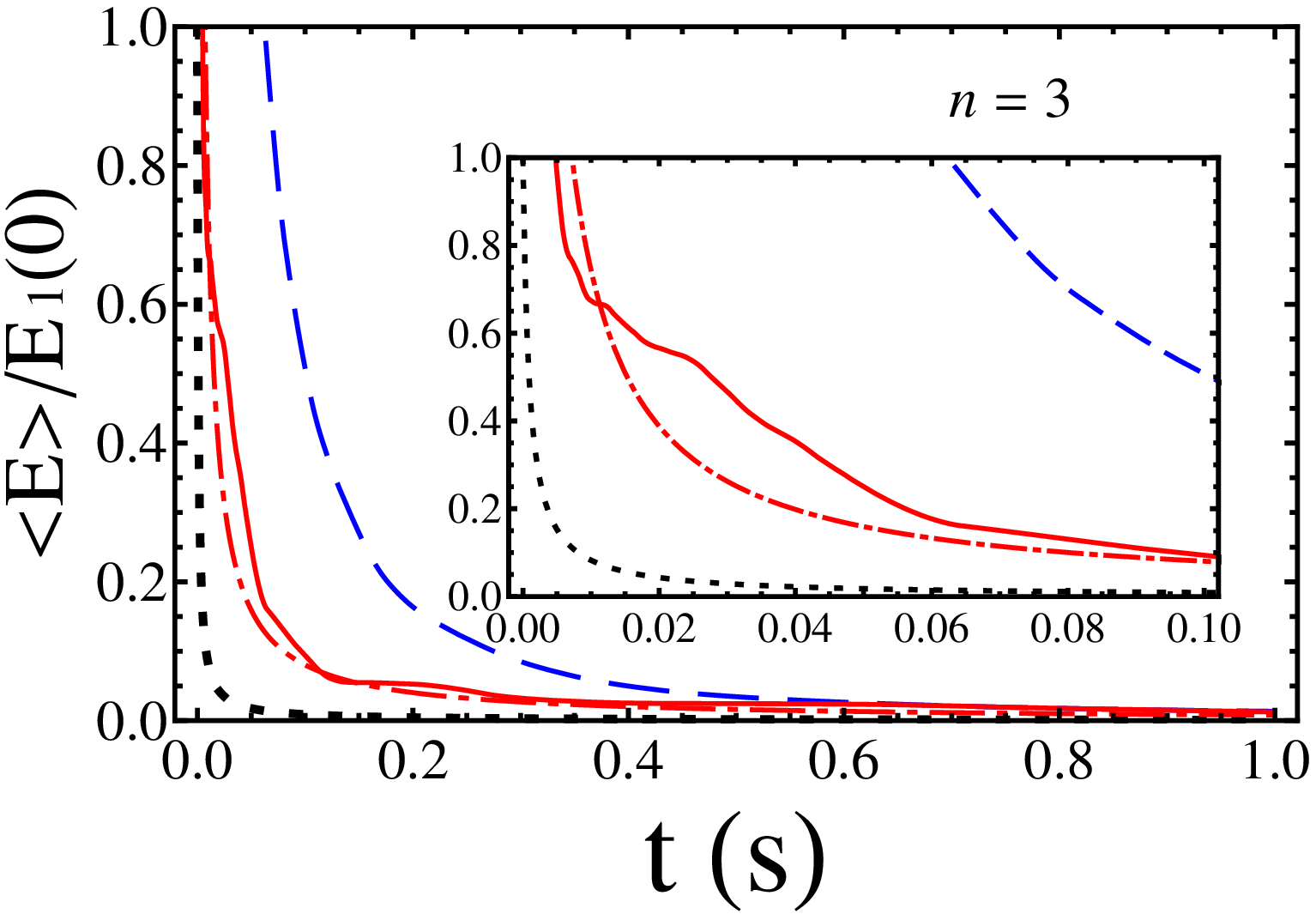}}
\end{center}
\caption{\label{fig4} (Color online) Dependence of the
average energy $\langle E \rangle$ on $t$ for different
initial states $u_n (0)$ with $n=1,2,3$:
square-root mirror (solid red line), linear mirror (dashed blue line);
$t_f=1$ s, $L_0=3$ $\mu$m, $L_f= 100$ $\mu$m.
The adiabatic energy $E_{n}(t)$ for the linear mirror is hardly
distinguishable from 
the dashed line, whereas for the square-root mirror it is depicted 
as a red dot-dashed curve. The black dotted curve is 
the adiabatic energy of the ground state $E_{1}(t)$ for the square-root mirror.}
\end{figure}
 
{\it iii) Condition of adiabaticity}. To maintain adiabaticity during the
expansion, the system should satisfy
\begin{equation}
\left|\frac{\langle u_k(t)|\frac{d}{dt} u_n(t)
\rangle}{(E_{k}(t)-E_n(t))/\hbar}\right|\ll 1
\end{equation}
for populated levels and their neighboring levels \cite{Schiff}.
Applying this to the 
expanding box for $k=2$ and $n=1$ at   
$t_f$, when the levels are at closest distance from each other, and neglecting
$L_0/L_f$ terms, we get
\begin{equation}
\frac{8}{9 \pi} \frac{m L_f^2}{\pi \hbar} \ll t_f
\end{equation}
or 
$t_f E_{adiab}(t_f) \gg \frac{4}{9} \hbar$, 
which is -because of $L_0 \ll L_f$- a stronger condition than the condition
(\ref{condstop}) for the square-root mirror. Earlier breakdown of adiabaticity 
for the linear mirror results, in accordance with the Berry-Klein asymptotics, in a
finite final energy, whereas, as we shall see, 
the square-root mirror provides efficient cooling 
as long as the stopping condition is satisfied, irrespective of adiabaticity.  

For sufficiently large $t_f$ and small $L_f$, the whole expansion can 
be adiabatic, even for the square-root case, as in Fig. \ref{fig2}b.
In adiabatic processes like the one depicted, 
the square-root mirror makes a faster approach to the final energy than the
linear one, as a consequence of the inequality $L_{square-root}(t)\geq L_{linear}(t)$.  

To determine how much shorter the expansion time $t_f$ could be,   
we compare in Fig. \ref{fig3}a for the atom initially in the ground state,
the final average energies  with respect to  $t_f$ with a fixed
$L_f$.
The square-root is always more efficient than the 
linear mirror. In particular at $t_f=0.3$ s a 1/100 reduction of the initial energy 
is achieved. Fig. \ref{fig2}a provides details of the evolution for that particular final time. The expansion for the linear mirror 
is adiabatic only up to some time ($\approx 0.015$ s), and in the final stage of the evolution the constant energy value predicted asymptotically by
Berry and Klein is reached.
On the other hand, the square-root mirror expansion is never adiabatic.
Efficient cooling starts at time $\tau_C \approx 0.02$ s, and this delay may be
understood classically since the square-root
mirror is quite fast at first so that the particles need some time to
catch up with the mirror.
In spite of the lack of adiabaticity, the stopping criterion
(\ref{condstop}) is well satisfied, so that the end result is a
very fast and efficient cooling.
Note that the minimal work principle is respected at all times. 

Finally, we compare the behavior of different
initial states. Fig. \ref{fig4} shows the average energy for different
initial states, $n=1,2,3$.
When the initial state is an excited state, 
the density matrix is not passive: this allows $\langle E \rangle < E_{adiab}$
to occur, a (justified) transient violation of the microscopic minimal
work principle, see the inset of Fig. \ref{fig4}b. 
Moreover one notes in Fig. \ref{fig4} that 
the difference between the curves for square-root and linear expansions 
for short $t$ increases with the excitation.

This suggests a good behavior of the 
square-root expansion also for simple systems such as a polarized Fermi
gas, or its (symmetrized) bosonic counterpart, the Tonks-Girardeau gas.
The average energy of a Tonks-Girardeau gas \cite{tg} with $N$ particle
is given by $\langle E \rangle_{TG} = \sum_{n=1}^N \langle E
\rangle_{n}$ where $\langle E \rangle_{n}$ is the average energy
of the single-particle solution with initial state $u_n$.
In Fig. \ref{fig3}b, we
compare the average energy of the square-root and the linear mirror
for a Tonks-Girardeau gas with $5$ particles
for different final times $t_f$. 

In conclusion, square-root in time box expansions offer advantages over
linear in time expansions, allowing faster and more efficient 
cooling.
While the breakdown of adiabaticity is generally
fatal for cooling with the linear expansion, this is not the case
for square-root expansion.   
The universal stopping and cooling properties of square-root expansions
are also in contrast to generic state-dependence of
variational methods.
We close noticing that the box studied can be realized
experimentally in an all-optical implementation \cite{Mark3}.
The effect of alternative trapping potentials and mirror trajectories
as well as further many-body effects will be reported elsewhere.

J. G. M. acknowledges M. Berry for useful comments.
J. G. M. and A. R. acknowledge the kind hospitality of the Max Planck
Institute for Complex Systems at Dresden.
We acknowledge funding by Projects No. GIU07/40,  FIS2006-10268-C03-01,
60806041, 08QA14030, 2007CG52, Juan de la Cierva Program, EU Integrated
Project QAP, EPSRC QIP-IRC, and the German Research Foundation (DFG). 
%

\end{document}